\documentclass[sigconf,screen]{acmart}
\usepackage[utf8]{inputenc}
\usepackage{subcaption}
\usepackage{siunitx}
\usepackage{wrapfig}
\usepackage{makecell}
\usepackage{verbatim}
\usepackage{xcolor}
\usepackage{colortbl}
\usepackage{url}
\usepackage{subcaption}
\usepackage{listings}
\usepackage{setspace}
\usepackage[ruled,vlined,linesnumbered]{algorithm2e}

\usepackage{enumitem}
\usepackage[noend]{algpseudocode}
\usepackage{float}

\usepackage{amsthm}

\algblock{Input}{EndInput}
\algblock{Output}{EndOutput}

\usepackage{soul}
\setstcolor{blue}

\usepackage{xparse}
\usepackage{mathtools}
\usepackage{multirow}
\usepackage{datapie}
\usepackage{adjustbox}
\usepackage[most]{tcolorbox}
\usepackage{pgf}
\usepackage{eqparbox}
\newdimen{\algindent}
\algnewcommand\LeftComment[2]{%
\hspace{#1\algindent}$\triangleright$ #2 \hfill %
}

\newcommand{\headercolorlong}{\rowcolor{gray!17}}
\newcommand{\highlighttext}[1]{\colorbox[rgb]{1.0, 0.9, 0.8}{\textcolor{black}{#1}}}

\newcommand{\modify}[1]{#1}

\newcommand{\benchtheorems}{1,720}
\newcommand{\benchprojects}{14}
\newcommand{\benchavgsteps}{16.1}
\newcommand{\benchmediansteps}{7.0}

\newcommand{\coq}{Rocq}

\newcommand{\proofstate}{state of the proof assistant}

\newcommand{\oursonbench}{651}
\newcommand{\rangoonbench}{481}
\newcommand{\palmonbench}{522}
\newcommand{\cobblestoneonbench}{527}
\newcommand{\selfrefineonbench}{503}
\newcommand{\selfrefineragonbench}{549}

\newcommand{\oursoncoqstoq}{484}
\newcommand{\rangooncoqstoq}{352}
\newcommand{\palmoncoqstoq}{415}
\newcommand{\cobblestoneoncoqstoq}{435}
\newcommand{\selfrefineoncoqstoq}{357}
\newcommand{\selfrefineragoncoqstoq}{410}
\newcommand{\oursfive}{196}
\newcommand{\oursfouro}{188}
\newcommand{\oursfouromini}{152}
\newcommand{\oursclaude}{185}
\newcommand{\oursdeepseek}{179}
\newcommand{\oursllama}{177}

\newcommand{\oursnorag}{622}
\newcommand{\oursnolemma}{590}
\newcommand{\oursbase}{519}
\newcommand{\oursfullstate}{625}

\newcommand{\oursonbenchp}{37.85\%}
\newcommand{\rangoonbenchp}{27.97\%}
\newcommand{\palmonbenchp}{30.35\%}
\newcommand{\cobblestoneonbenchp}{30.64\%}
\newcommand{\selfrefineonbenchp}{29.24\%}
\newcommand{\selfrefineragonbenchp}{31.92\%}

\newcommand{\oursoncoqstoqp}{41.33\%}
\newcommand{\rangooncoqstoqp}{30.06\%}
\newcommand{\palmoncoqstoqp}{35.44\%}
\newcommand{\cobblestoneoncoqstoqp}{37.15\%}
\newcommand{\selfrefineoncoqstoqp}{30.49\%}
\newcommand{\selfrefineragoncoqstoqp}{35.01\%}
\newcommand{\oursoutselfrefineragonbench}{18.58\%}
\newcommand{\oursoutpalmonbench}{24.71\%}
\newcommand{\oursoutcobblestoneonbench}{23.53\%}
\newcommand{\oursoutrangoonbench}{35.34\%}

\newcommand{\oursoutpalmoncoqstoq}{16.63\%}
\newcommand{\oursoutcobblestoneoncoqstoq}{11.26\%}
\newcommand{\oursoutrangooncoqstoq}{37.50\%}
\newcommand{\oursfivep}{39.20\%}
\newcommand{\oursfourop}{37.60\%}
\newcommand{\oursfourominip}{30.40\%}
\newcommand{\oursclaudep}{37.00\%}
\newcommand{\oursdeepseekp}{35.80\%}
\newcommand{\oursllamap}{35.40\%}

\newcommand{\oursnoragp}{36.16\%}
\newcommand{\oursnolemmap}{34.30\%}
\newcommand{\oursbasep}{30.17\%}
\newcommand{\oursfullstatep}{36.34\%}
\newcommand{\oursoutnorag}{4.66\%}
\newcommand{\oursoutnolemma}{10.34\%}

\newcommand{\oursuniquebench}{91}
\newcommand{\baselineuniquebench}{133}
\newcommand{\oursuniquecoqstoq}{54}
\newcommand{\baselineuniquecoqstoq}{61}

\newcommand{\oursuniquebenchp}{5.29\%}
\newcommand{\baselineuniquebenchp}{7.73\%}
\newcommand{\oursuniquecoqstoqp}{4.61\%}
\newcommand{\baselineuniquecoqstoqp}{5.21\%}
\newcommand{\name}{\textsc{Adapt}}
\newcommand{\bench}{\textsc{CoqDev}}
\newcommand{\coqstoq}{\textsc{CoqStoq}}

\newcommand{\palm}{\textsc{PALM}}
\newcommand{\rango}{\textsc{Rango}}
\newcommand{\cobblestone}{\textsc{Cobblestone}}
\newcommand{\selfrefine}{\textsc{Self-Refine}}
\newcommand{\selfrefinerag}{\textsc{Self-Refine+RAG}}

\newcommand{\namenorag}{\textit{\name{}\_no\_enrich}}
\newcommand{\namenolemma}{\textit{\name{}\_no\_lemma}}
\newcommand{\namebase}{\textit{\name{}\_base}}
\newcommand{\namefullstate}{\textit{\name{}\_full\_state}}

\newcommand{\gptfive}{GPT-5.2}
\newcommand{\gptfouro}{GPT-4o}
\newcommand{\gptfouromini}{GPT-4o-mini}
\newcommand{\claude}{Claude-3.7-Sonnet}
\newcommand{\deepseek}{DeepSeek-V3}
\newcommand{\llama}{Llama-4-Maverick-17B-128E}

\newcommand{\nameruledecision}{\textsc{Adapt}\textsubscript{\textit{rule}}}
\newcommand{\namemldecision}{\textsc{Adapt}\textsubscript{\textit{ml}}}
\newcommand{\namerandomdecision}{\textsc{Adapt}\textsubscript{\textit{random}}}
\newcommand{\nameseqdecision}{\textsc{Adapt}\textsubscript{\textit{seq}}}

\newcommand{\coqhammer}{\textit{CoqHammer}}

\lstdefinestyle{smallCoq}{
    style=coq,
    basicstyle=\ttfamily\footnotesize,
    breaklines=true,
    frame=none,
    xleftmargin=0pt,
    aboveskip=0pt,
    lineskip=-2pt,
    belowskip=0pt
  }

\AtBeginDocument{
  }

\setcopyright{cc}
\setcctype{by}
\acmDOI{10.1145/3832783.3834382}
\acmYear{2026}
\copyrightyear{2026}
\acmISBN{979-8-4007-2882-2/2026/10}
\acmConference[ASE '26]{Proceedings of the 41st IEEE/ACM International Conference on Automated Software Engineering}{October 12--16, 2026}{Munich, Germany}
\acmBooktitle{Proceedings of the 41st IEEE/ACM International Conference on Automated Software Engineering (ASE '26), October 12--16, 2026, Munich, Germany}
\acmSubmissionID{ase26main-p1291-p}
\received{2026-03-26}
\received[accepted]{2026-06-18}

\begin{document}

\title{Adaptive Proof Refinement with LLM-Guided Strategy Selection}

\definecolor{lightgray}{gray}{0.97}
\definecolor{keywordcolor}{rgb}{0.13, 0.13, 1} 
\definecolor{keywordcolor2}{rgb}{0.5,0,0.5}

\lstdefinelanguage{Coq}{
  morekeywords={Type, Prop, Definition, Lemma, Theorem, Proof, Qed, Fixpoint, Inductive, Notation},
  sensitive=true, 
}

\lstdefinestyle{coqline}{
    breakatwhitespace=false,
    breaklines=true,
    captionpos=t,
    keepspaces=true,
    basicstyle=\ttfamily\footnotesize,
    showspaces=false,
    showstringspaces=false,
    showtabs=false,
    tabsize=2,
    language=Coq,
    keywordstyle=\color{keywordcolor},
    breaklines=true,
    frame=tb,
    framesep=1pt,
    framerule=0.5pt,
    rulecolor=\color{black},
    aboveskip=0pt,
    belowskip=0pt,
    escapeinside={(*@}{@*)},
    numbersep=0pt,
    numbers=left,
    numberstyle=\tiny,
  }

\lstdefinestyle{coq}{
    breakatwhitespace=false,
    breaklines=true,
    captionpos=t,
    keepspaces=true,
    basicstyle=\ttfamily\footnotesize,
    showspaces=false,
    showstringspaces=false,
    showtabs=false,
    tabsize=2,
    language=Coq,
    keywordstyle=\color{keywordcolor},
    breaklines=true,
    frame=single,
    framesep=1pt,
    framerule=0.5pt,
    rulecolor=\color{black},
    aboveskip=0pt,
    belowskip=0pt,
    escapeinside={(*@}{@*)}
  }


\author{Minghai Lu}
\orcid{0009-0001-0136-3204}
\affiliation{%
  \institution{Purdue University}
  \city{West Lafayette}
  \country{USA}
}
\email{lu1074@purdue.edu}

\author{Zhe Zhou}
\authornote{Corresponding author.}
\orcid{https://orcid.org/0000-0003-3900-7501}
\affiliation{%
  \institution{Purdue University}
  \city{West Lafayette}
  \country{USA}
}
\email{zhou956@purdue.edu}

\author{Danning Xie}
\authornote{Work done while at Purdue University.}
\orcid{https://orcid.org/0000-0002-4359-4625}
\affiliation{%
  \institution{Meta}
  \city{Menlo Park}
  \country{USA}
}
\email{dxie@meta.com}

\author{Songlin Jia}
\orcid{https://orcid.org/0009-0008-2526-0438}
\affiliation{%
  \institution{Purdue University}
  \city{West Lafayette}
  \country{USA}
}
\email{jia137@purdue.edu}

\author{Benjamin Delaware}
\orcid{0000-0002-1016-6261}
\affiliation{%
  \institution{Purdue University}
  \city{West Lafayette}
  \country{USA}
}
\email{bendy@purdue.edu}

\author{Tianyi Zhang}
\orcid{0000-0002-5468-9347}
\affiliation{%
  \institution{Purdue University}
  \city{West Lafayette}
  \country{USA}
}
\email{tianyi@purdue.edu}


\begin{abstract}
  Formal verification via theorem proving enables rigorous proofs of
  software correctness, but it is difficult to scale due to the
  significant manual effort and expertise required. While Large
  Language Models (LLMs) have shown potential in automating proof
  generation, they frequently produce incorrect proofs on the first
  attempt that require iterative refinement to fix. However, most
  existing approaches employ fixed refinement strategies and cannot
  dynamically choose an effective strategy, which limits their
  performance. To overcome this limitation, we introduce \name{}, a
  novel proof refinement framework that leverages an LLM-guided
  decision-maker to dynamically select a suitable refinement strategy
  according to the \proofstate{} and available context of an incorrect
  proof. We evaluate \name{} on two benchmark suites against five
  existing methods and find that it significantly outperforms the best
  baseline on both by proving \oursoutcobblestoneoncoqstoq{} and
  \oursoutselfrefineragonbench{} more theorems,
  respectively. Furthermore, we demonstrate \name{}'s generalizability
  by evaluating it across six different LLMs. We also conduct ablation
  studies to measure the contribution of each component and compare
  the trade-offs of alternative decision-maker designs.
\end{abstract}


\maketitle

\captionsetup[figure]{font=bf,skip=6pt}
\captionsetup[table]{font=bf,skip=6pt}
\newcommand{\distance}{8pt}
\setlength{\textfloatsep}{6pt}
\setlength{\floatsep}{\distance}
\setlength{\dblfloatsep}{\distance} 

\section{Introduction}
Interactive theorem provers (ITPs) such as \coq{}~\cite{coq} and
Lean~\cite{lean} are powerful tools for formally reasoning about
complex software systems. These tools have been used to mechanically
verify deep correctness properties of safety-critical applications,
including compilers~\cite{CompCert-ERTS-2018}, distributed
systems~\cite{verdi}, and OS kernels~\cite{klein2009sel4, CertiKOS}.
These successes have come at the cost of significant human effort and
expertise, however, requiring developers to interactively guide the
proof assistant step-by-step. The proof of correctness for the
CompCert verified C compiler, for example, consists of 100,000
lines of proof scripts~\cite{CompCert-ERTS-2018} that took 6
person-years to develop.

Motivated by these costs, recent years have seen increasing interest
in using Large Language Models (LLMs) to automate the development of
proofs in ITPs~\cite{thompson2025rango, first2023baldur,
  yang2024leandojo,
  wang2023lego, cao2025informal, liu2025proofaug}.
Many existing approaches adopt an iterative refinement paradigm---they
use an LLM to generate an initial proof attempt and then incrementally
repair that proof~\cite{lu2024proof, first2023baldur, wang2023lego,
  liu2025proofaug, an-etal-2024-learn,
  yang2025autoverus}.
Yet they either rely on a single refinement strategy to repair every
proof error~\cite{an-etal-2024-learn, first2023baldur,
  wang2023lego,liu2025proofaug} or use a set of hard-coded rules to
select a repair strategy based solely on error
messages~\cite{lu2024proof, yang2025autoverus}. This stands in
contrast to how human experts fix broken proofs: rather than relying
on a single error message, developers comprehensively analyze the
entire context of a failed proof attempt to decide between different
repair strategies.  For example, they may introduce new helper lemmas
when existing ones are insufficient, draw inspiration from similar
proofs and reuse relevant lemmas, or iteratively repair proofs based
on ITP feedback. This sort of strategic decision-making is
particularly important in the ITP setting, where a wrong choice may
send a proof down an unproductive path, compounding the mistake by
significantly increasing proof effort or eventually leading to
outright failure.

\begin{figure}[thb]
\vspace{-5pt}
  \includegraphics[width=0.5\textwidth]{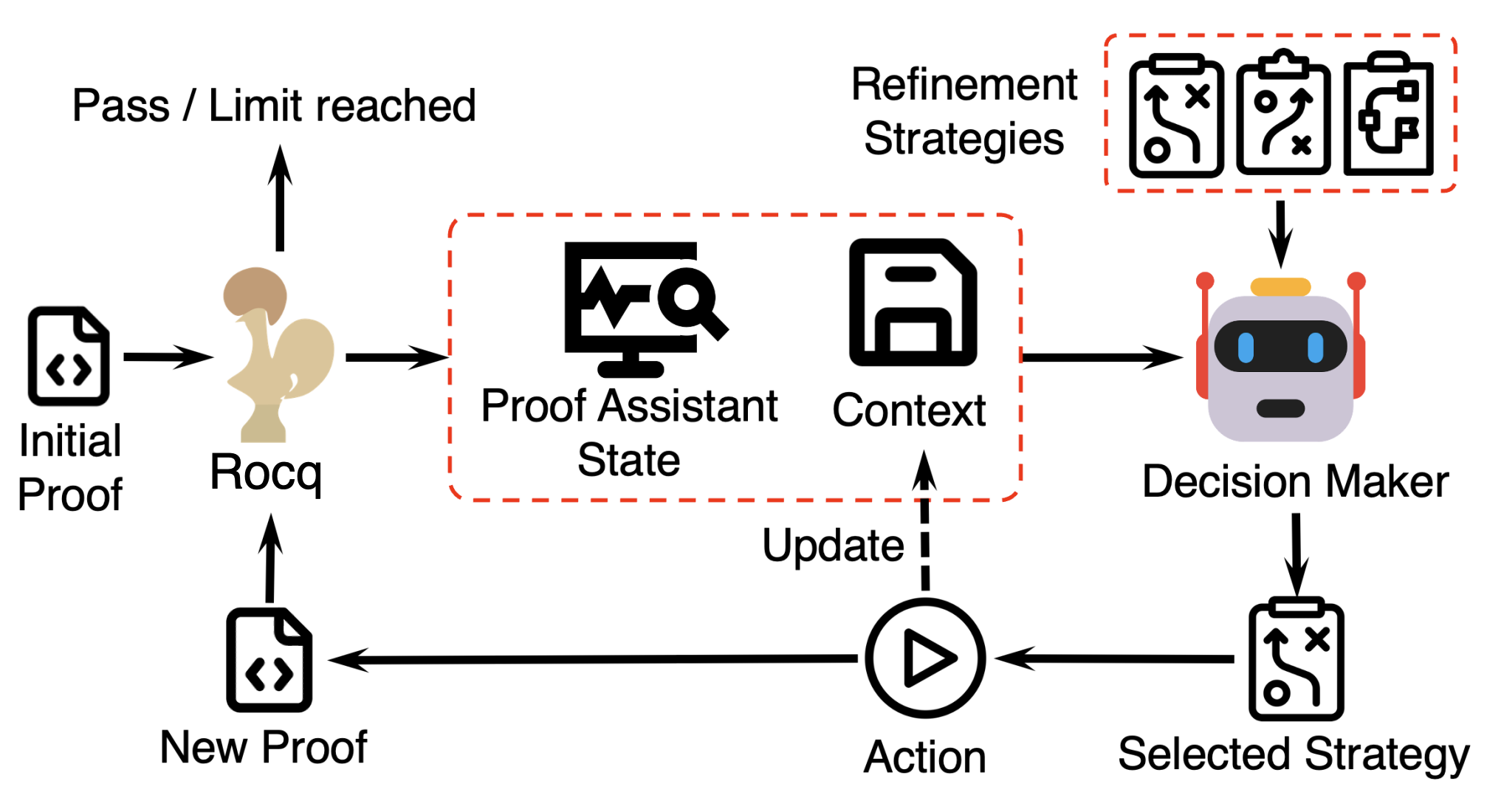}
  \caption{\label{arch} Overview of \name{}.}
\end{figure}

To address this limitation in prior work, we propose \name{}, an
adaptive proof refinement framework that emulates the decision-making
process of human experts. \name{} employs an LLM-guided decision-maker
to analyze the current \proofstate{} and available context, then
dynamically selects an appropriate strategy. As illustrated in
Figure~\ref{arch}, when a proof attempt fails, the decision-maker
employs one of three strategies: (1) proposing new helper lemmas, (2)
searching for relevant lemmas and proofs to guide the refinement, or
(3) regenerating the proof. This iterative process allows \name{} to
flexibly adapt its approach at each step, enabling it to effectively
prove complex theorems.

Our evaluation of \name{} against three state-of-the-art methods,
\cobblestone{}~\cite{kasibatla2025cobblestone},
\palm{}~\cite{lu2024proof} and \rango{}~\cite{thompson2025rango}, and
two alternative refinement designs demonstrates its effectiveness. On
the \coqstoq{} benchmark~\cite{thompson2025rango}, \name{} proves
\oursoncoqstoqp{} of 1,171 theorems, outperforming the strongest
baseline by \oursoutcobblestoneoncoqstoq{}. This lead increases to
\oursoutselfrefineragonbench{} on \bench{}, a new, more challenging
benchmark comprised of \benchtheorems{} theorems that we have
developed. Further experiments confirm the generalizability of \name{}
across six LLMs and, through ablation studies, validate the
contribution of each refinement strategy and the LLM-guided
decision-maker.

In summary, we make the following contributions:
\begin{enumerate}
\item We propose \name{}, an adaptive proof refinement framework that
  dynamically selects from different refinement strategies based on
  the proof state and available context.
\item We construct \bench{}, a new, more challenging benchmark mined
  from real-world \coq{} commit histories that model the incremental
  development process.
\item We conduct a comprehensive evaluation of \name{} on
  \bench{} and the most recent  \coqstoq{} benchmark, demonstrating that
  \name{} achieves significant improvements over state-of-the-art
  techniques.
\end{enumerate}

\section{Background and Motivation}
\label{sec:background}
\begin{figure}[t]
\centering
\begin{subfigure}[h]{0.48\textwidth}
\lstset{style=coqline}
\begin{lstlisting}
 (*@\textcolor{gray}{(* Definitions *)}@*)
 Inductive nat : Set := (*@\textcolor{gray}{(* Datatype of natural numbers *)}@*)
   | O : nat
   | S : nat -> nat.
 Notation "n + 1" := (S n).

 Fixpoint leb (n m : nat) : bool := (*@\textcolor{gray}{(* Comparison function *)}@*)
   match n, m with
     | O, _ => true
     | _, O => false
     | n + 1, m + 1 => leb n m
   end.

 Inductive le: nat -> nat -> Prop := (*@\textcolor{gray}{(* Comparison relation *)}@*)
   | le_n (n : nat) : le n n
   | le_S (n m : nat) (H : le n m) :
     le n (S m).
 Notation "n <= m" := (le n m).

 (*@\textcolor{gray}{(* Theorems *)}@*)
 Lemma O_le_n : forall n, O <= n.
 Proof.
   intros n. induction n.
   - apply le_n.
   - apply (le_S O n IHn).
 Qed.

 Lemma n_le_m_Sn_le_Sm :
   forall n m, n <= m ->  (n + 1) <= (m + 1).
 Proof. (*@\textcolor{gray}{(* Proof omitted *)}@*) Qed.

 Theorem leb_complete : forall n m,
   leb n m = true -> n <= m.
 Proof.
   intros n.
   induction n.  (*@\textcolor{gray}{(* Induction on n *)}@*)
   - intros.     (*@\textcolor{gray}{(* Base case *)}@*)
     apply O_le_n.
   - intros.     (*@\textcolor{gray}{(* Inductive case *)}@*)
     destruct m. (*@\textcolor{gray}{(* Case analysis on m *)}@*)
     + discriminate.
     + apply n_le_m_Sn_le_Sm. (*@\textcolor{gray}{(* Use lemma *)}@*)
       simpl in H. apply IHn in H. apply H.
 Qed.
\end{lstlisting}
\end{subfigure}
\caption{An example \coq{} development that includes a datatype
  definition (\texttt{nat}), a recursive function (\texttt{leb}), and
  an inductive relation (\texttt{le}), two lemmas and a theorem.}
\label{example:context}
\end{figure}

\begin{figure}[htb]
\centering
\begin{subfigure}[b]{\linewidth}
\lstset{style=coq}
\begin{lstlisting}
m: nat
H: leb 0 m = true
================================================
0 <= m
\end{lstlisting}
\caption{Proof state before line 38 in Figure~\ref{example:context}.}
\vspace{10pt}
\label{example:state_0}
\end{subfigure}

\begin{subfigure}[b]{\linewidth}
\lstset{style=coq}
\begin{lstlisting}
n, m: nat
IHn: forall m: nat, leb n m = true -> n <= m
m: nat
H: leb (n + 1) (m + 1) = true
================================================
n + 1 <= m + 1
\end{lstlisting}
\caption{Proof state before line 42 in Figure~\ref{example:context}.}
\vspace{10pt}
\label{example:state_1}
\end{subfigure}
\begin{subfigure}[b]{\linewidth}
\lstset{style=coq}
\begin{lstlisting}
n: nat
IHn: forall m: nat, n <= m -> leb n m = true
m: nat
H: n + 1 <= m + 1
================================================
leb (n + 1) (m + 1) = true
\end{lstlisting}
\caption{Proof state before line 52 in Figure~\ref{example:proof}.}
\label{example:state_2}
\end{subfigure}

\caption{Intermediate proof states. (a) shows the goal for the base case; (b) shows the inductive step requiring a helper lemma; (c) shows a state during the proof of \texttt{leb\_correct}.}
\label{fig:proof_states}
\end{figure}

\begin{figure}[t]
\vspace{10pt}
\centering
\begin{subfigure}[h]{0.45\textwidth}
\lstset{style=coqline}
\begin{lstlisting}[firstnumber=45]
 Theorem leb_correct :
    forall n m, n <= m -> leb n m = true.
 Proof.
   intros n. induction n.          (*@\textcolor{gray}{(* Induction on n *)}@*)
   - intros. simpl. reflexivity.   (*@\textcolor{gray}{(* Base case *)}@*)
   - intros. destruct m.           (*@\textcolor{gray}{(* Inductive case *)}@*)
     + inversion H.
     + apply Sn_le_Sm_n_le_m in H. (*@\textcolor{gray}{(* Use lemma *)}@*)
       simpl. apply IHn in H. apply H.
 Qed.
\end{lstlisting}
\caption{\label{example:proof} A \coq{} theorem stating the correctness of the \texttt{leb} function with respect to the \texttt{le} relation, and its proof script.}
\end{subfigure}

\vspace{10px}

\begin{subfigure}[h]{0.45\textwidth}
\lstset{style=coqline}
\begin{lstlisting}[firstnumber=55]
 Lemma Sn_le_Sm_n_le_m :
    forall n m, (n + 1) <= (m + 1) -> n <= m.
 Proof. (*@\textcolor{gray}{(* Proof omitted *)}@*) Qed.
\end{lstlisting}
\caption{\label{example:lemma} A helper lemma used in the proof of
  \texttt{leb\_correct} above.}
\end{subfigure}

\caption{\label{example} Example of proof construction by searching and reusing existing proofs.}
\label{fig:enter-label}
\end{figure}

\noindent To motivate our ideas, Figure~\ref{example:context} presents
a self-contained example of a proof development in
\coq{}~\cite{coq}---a widely used theorem prover for software
verification and the primary ITP considered in this work. We start
with a brief overview of the proof process in \coq{} and then
introduce the problem setting of adaptive proof refinement.

\paragraph{Definitions and Theorems} As is common in \coq{}, the example in Figure~\ref{example:context} begins with the \textit{\textbf{definition}} of data structures, functions, and inductive relations. Lines 1--4 define Peano natural numbers (\texttt{nat}) with a base constructor (\texttt{O}) and an inductive successor constructor (\texttt{S}), where \texttt{S n} returns the number that is one more than \texttt{n}. The file then defines a boolean function \texttt{leb} that recursively checks if one number is less than or equal to another, and an inductive relation \texttt{le} that provides a formal definition of the same concept.

After the definitions, \textit{\textbf{theorems}} are stated and proven. They start with the keyword \texttt{Theorem} or \texttt{Lemma}, which are functionally identical but used by convention to distinguish key results from helper proofs. The main theorem, \texttt{leb\_complete} (line 32), states that if \texttt{leb n m} is \texttt{true}, then \texttt{n} is less than or equal to \texttt{m}. To prove this, two helper lemmas are required: \texttt{O\_le\_n}, which proves that \texttt{O} is less than or equal to any natural number, and \texttt{n\_le\_m\_Sn\_le\_Sm}, which shows that if \texttt{n <= m}, then \texttt{n + 1 <= m + 1}.

\paragraph{Proof State and Context} Given a theorem statement, \coq{}
enters an interactive proof mode and displays the current
\textit{\textbf{proof state}}. This state includes the \textit{goal}
(i.e., the obligation that the user needs to prove) and the
\textit{\textbf{local context}} (i.e., any local variables and
hypotheses used in that goal). The user applies
\textit{tactics} to transform this state by, e.g., simplifying the
goal or introducing new hypotheses. \coq{} then updates the state or
reports an error when tactic application is invalid. The proof of
\texttt{leb\_complete} uses the \texttt{induction} tactic on
\texttt{n} (line 36) and case analysis (\texttt{destruct}) on
\texttt{m} (line 40). Figure~\ref{example:state_0} shows the proof
state before the tactic on line 38 is applied; the context is above
the double line and the goal is below.

\paragraph{Proof Failure} \coq{} reports an error for invalid tactic
applications. For instance, under the proof state in
Figure~\ref{example:state_1} derived from line 42 of
Figure~\ref{example:context}, the \textit{erroneous tactic}
\texttt{apply n\_le\_m\_Sn\_le\_Sm in H} leads to a proof failure. It
attempts to perform forward reasoning by applying the lemma to the
hypothesis \texttt{H}, but fails with the \textit{error message}:
\textit{``Unable to apply lemma of type "forall n m : nat, n <= m -> n
  + 1 <= m + 1" on hypothesis of type "leb (n + 1) (m + 1) = true"''}.

\paragraph{Project Management and Refinement Strategies} To manage
complexity, formal proofs are often organized into a \textbf{\em proof
  project} with multiple files and imported libraries, similar to a
software project. When writing a proof, developers use not only the
local context but also a \textit{\textbf{global context}} of
definitions and theorems from other files and imported libraries. This
leads to several refinement strategies. One common strategy is to
search the current project for helpful lemmas. For example, to make
progress from the state in Figure~\ref{example:state_1}, the proven
lemma \texttt{n\_le\_m\_Sn\_le\_Sm} is applied to the goal,
simplifying it to \texttt{n <= m}. When no suitable existing lemma is
available, a developer may need to invent a new one, a strategy known
as \textit{\textbf{lemma discovery}}~\cite{alhessi2025lemmanaid,
  kurashige2024cclemma, sivaraman2022data, johansson2019lemma}.

Similar to how developers search and reuse existing code to write new code~\cite{sadowski2015developers}, existing proofs can also serve as a source of inspiration in proof construction.
For example, suppose a developer, after proving everything in
Figure~\ref{example:context}, now wants to prove \texttt{leb\_correct}
(Figure~\ref{example:proof}). The statement of \texttt{leb\_correct}
is structurally similar to \texttt{leb\_complete}, suggesting a
similar proof pattern (induction on \texttt{n}, case analysis on
\texttt{m}). Of course, na\"ively copying the proof of
\texttt{leb\_complete} and reusing it does not work, since the helper
lemmas used in \texttt{leb\_complete} do not directly apply to \texttt{leb\_correct}. An additional refinement strategy is required: the developer can draw inspiration from them to propose and prove a new, necessary lemma: \texttt{Sn\_le\_Sm\_n\_le\_m} (Figure~\ref{example:lemma}).

This example highlights the importance of \emph{adaptive proof
  refinement that integrates multiple proof refinement strategies}
according to the available context (i.e., local and global context) as
well as the current \proofstate{} (e.g., partial proof, erroneous
tactic, error message, and proof state). A tool that only retrieves
existing lemmas would fail because the necessary lemma
\texttt{Sn\_le\_Sm\_n\_le\_m} does not exist in the
\texttt{leb\_correct} example. Conversely, an approach focusing only
on lemma discovery would have to generate both the new lemmas and
proofs from scratch, which is difficult and costly without the
guidance of existing lemmas and proofs. The core challenge, therefore,
is to create a system that can dynamically assess the available
context and current \proofstate{} to select the most promising
strategy, like a human expert.

\section{Approach}

Given a theorem statement to prove, \name{} first generates an initial proof and checks it using \coq{}. If the proof fails, \name{} captures the \textit{\proofstate{}}, including the erroneous tactic, the proof state at the point of failure, the error message from \coq{}, and the partial proof preceding the error. It then initializes a working context by retrieving essential information from the global context, including relevant lemmas and similar proofs in the repository.

\name{} then iteratively refines the failed proof attempt. In each iteration, the decision-maker analyzes the \proofstate{} and working context to select and apply a strategy, including lemma discovery, context enrichment, and regeneration. \modify{If the attempt fails, \name{} proceeds to the next iteration using the updated \proofstate{} and working context.} This process continues until a successful proof is found or an iteration limit is reached. The rest of this section elaborates on each of these components.

\subsection{Initial Proof Generation}
\label{sec:initial_proof}
\name{}'s iterative process begins with an initial proof attempt. \name{} is agnostic to the underlying proof generator, and can freely use, e.g., LLM-based~\cite{thompson2025rango, lu2024proof, wang2023lego, first2023baldur}, deep learning-based~\cite{yang2019learning, sanchez2023passport, first2020tactok, first2022diversity}, or symbolic approaches~\cite{bundy2002critique}. In this work, we use the retrieval-augmented generation method proposed by Lu et al.~\cite{lu2024proof} to generate the initial proof, since it is a recent method with the source code publicly available. \modify{Specifically, this approach first identifies candidate theorems and lemmas using TF-IDF and KNN, then reranks them with BM25~\cite{bm25}.} Figure~\ref{prompt_inital} shows the prompt template used to generate the initial proof.\footnote{\modify{Coq was recently renamed to Rocq. Our experiments were conducted before the name change; our prompt templates retain the name ``Coq'', which was used in our experiments.}} The prompt includes the target theorem statement to be proven, all definitions appearing in the statement, and the top K relevant theorems and lemmas. \modify{When fewer than K candidates are available, all are included.} In this work, we set K to 10 to avoid exceeding the context window limit.

\begin{figure}[thb]
\footnotesize
\begin{tabular}{>{\raggedright\arraybackslash\tt}p{0.45\textwidth}}
\toprule
\headercolorlong
\textbf{INSTRUCTIONS}\\
You will be given a theorem statement written in Coq, with related definitions, theorems, and lemmas. Your job is to write a proof. You should first analyze how to prove it based on the definitions, and which theorems and lemmas can be used. Leverage as many theorems and lemmas as possible to facilitate your proof. \\

\headercolorlong
\textbf{INPUT}\\
- Theorem Statement: \highlighttext{<theorem>} \\
- Definitions: \highlighttext{<definitions>} \\
- Theorems and Lemmas: \highlighttext{<theorems and lemmas>} \\
\bottomrule
\end{tabular}
\caption{\label{prompt_inital} Prompt template for initial proof generation.}
\end{figure}

\subsection{Context Initialization}
\label{sec:context_init}

If the initial proof attempt fails, \name{} collects several pieces of information to help guide its iterative refinement loop.  It first captures the \proofstate{}, which includes the following:
\begin{itemize}
    \item \textbf{Erroneous tactic:} the specific tactic that caused the error.
    \item \textbf{Error message:} \coq{}'s explanation of why the tactic failed.
    \item \textbf{Partial proof:} the proof script before the erroneous tactic.
    \item \textbf{Stuck proof state:} the proof state resulting from the execution of the partial proof, which includes the remaining goals to prove and any local variables and hypotheses that are available, as defined in Section~\ref{sec:background}.
\end{itemize}

The combination of the erroneous tactic, its error message, and the stuck proof state reveals the immediate cause of failure, while the stuck proof state and the partial proof indicate the overall progress of the current proof. For instance, a simple syntax error in a nearly completed proof could be fixed by regeneration, whereas a failure near the beginning could signal a more fundamental issue with the overall proof strategy. This comprehensive information reflects the feedback a human developer receives and enables the decision-maker to examine different aspects of the failure and select a proper refinement strategy. 

In addition to the \proofstate{}, \name{}'s  refinement loop maintains a working context that guides its decision-maker and refinement strategies. \name{} initializes this context with the statements of theorems and lemmas retrieved during the initial proof generation, and the proof script of a similar theorem from the global context. Due to the context window limit of the LLM, we only include one proof script as an example. This proof provides a potential proof structure to guide subsequent proof attempts.
Following prior work~\cite{thompson2025rango}, we select a similar theorem using BM25~\cite{bm25}, which compares the textual similarity between the statement of the target theorem and the statements of other theorems in the global context. \modify{We select the theorem with the highest similarity score and include its proof, without setting a minimum similarity threshold.} This initialized context is then provided to the decision-maker to select the refinement strategy.

\begin{figure}[thb]
\footnotesize
\begin{tabular}{>{\raggedright\arraybackslash\tt}p{0.45\textwidth}}
\toprule
\headercolorlong
\textbf{INSTRUCTIONS} \\
We are proving a theorem in Coq but get stuck in a proof state, you need to select a refinement strategy to fix it. You will be provided with the theorem, and the diagnostic information, including the stuck proof state, the erroneous tactic that causes the error, the error message from Coq and the partial proof preceding the erroneous tactic. Additional supporting context, including related definitions and lemmas, and a similar theorem with its proof are also provided. \\

\headercolorlong
\textbf{INPUT} \\
- Theorem Statement: \highlighttext{<theorem>} \\
- Definitions: \highlighttext{<definitions>} \\
- Stuck Proof State: \highlighttext{<proof state>} \\
- Erroneous Tactic: \highlighttext{<erroneous tactic>} \\
- Error Message: \highlighttext{<error message>} \\
- Partial Proof: \highlighttext{<partial proof>} \\
- Theorems and Lemmas: \highlighttext{<theorems and lemmas>} \\
- Similar Proof: \highlighttext{<similar proof>} \\

\headercolorlong
\textbf{OUTPUT} \\
Analyze the provided information, then choose one of the following strategies: \\
\textbf{1. Strategy: Lemma Discovery} \\
Description: Propose new lemmas or refine existing ones to help the proof. \\
\highlighttext{Output Format:`Lemma Discovery; [name\_1, name\_2 ...]'}. This strategy always proposes new lemmas. Optionally, provide the names of existing lemmas or theorems to refine. Provide an empty list `[]' to only propose new lemmas. \\
\textbf{2. Strategy: Context Enrichment} \\
Description: Retrieve more relevant definitions, lemmas, theorems, and proofs from the project and imported libraries. \\
\highlighttext{Output Format:`Context Enrichment; [keyword\_1, keyword\_2,...]'}\\where you list keywords to search for, such as names of definitions or lemmas. Lemmas with the listed keywords in the statements will be retrieved. \\
\textbf{3. Strategy: Regeneration} \\
Description: Directly regenerate the proof if the current context seems sufficient or the previous proof is unpromising. \\
\highlighttext{Output Format:`Regeneration'}. \\
\bottomrule
\end{tabular}
\caption{\label{prompt_decision} Prompt template for decision-making.}
\end{figure}

\subsection{Decision-Maker}
\label{sec:decision_maker}
\name{} leverages the reasoning capabilities of LLMs to decide which refinement strategy to take based on both the current project and \proofstate{} described in Section~\ref{sec:context_init}. Figure~\ref{prompt_decision} shows the prompt template for decision making. Specifically, \name{} provides the LLM with the target theorem statement and the definitions that appear in it, the \proofstate{}, and the working context. The prompt then instructs the LLM to analyze the failure reasons and assess the sufficiency of the current context for solving the proof goal. Based on the analysis, the LLM is asked to select one of three predefined strategies. We do not claim that these three strategies are complete, as other useful refinement strategies may exist. Our goal is to demonstrate the effectiveness of adaptively selecting from different refinement strategies, rather than to propose all possible strategies.

\begin{enumerate}
    \item \textbf{Lemma discovery}: When a proof goal is too complex to be solved directly with the available context, human developers often propose new helper lemmas to decompose the proof into manageable sub-tasks. This process, known as \textit{lemma discovery}, has been proven to be effective when leveraging LLMs to propose new lemmas~\cite{wang2023lego, alhessi2025lemmanaid}. Inspired by this, we integrate the lemma discovery strategy to generate lemmas to decompose complex proof goals. As shown in Figure~\ref{prompt_decision}, when selecting this strategy, the decision-maker is also prompted to propose a set of existing theorems and lemmas that can be potentially adapted to generate a new lemma.
    \item \textbf{Context enrichment}: This strategy is designed to address failures caused by an insufficient working context by searching for additional definitions,  lemmas, theorems, and proofs.
    The decision-maker is also prompted to provide a set of keywords to guide the following context enrichment process, such as the names of definitions or lemmas present in the stuck proof state.
    \item \textbf{Regeneration}: This strategy is employed to either repair simple errors, such as a syntax mistake, or to try a new proof approach when the previous proof contains significant errors, such as applying induction on wrong variables or failing to generalize the induction hypothesis correctly.
\end{enumerate}

Alternative decision-maker designs offer different trade-offs in terms of performance and cost. Section~\ref{sec:decision_strategy} explores decision-makers based on heuristic rules and machine learning (ML) models. We describe each refinement strategy in the following sections. 

\subsection{Lemma Discovery Strategy}
\label{sec:lemma}

Lemma discovery takes as input the theorem statement, the stuck proof state, and the working context. The stuck proof state can reveal the specific difficulty encountered by the proof assistant, which helps \name{} generate lemmas tailored to the specific obstacle. For example, the proof state in Figure~\ref{example:state_2} suggests that a lemma establishing \texttt{(n + 1) <= (m + 1) -> n <= m} could be helpful, as it bridges the hypotheses \texttt{H} and \texttt{IHn}. Besides proposing new lemmas directly, \name{} also refines existing lemmas selected by the decision-maker (described in Section~\ref{sec:decision_maker}) to maximize the chance of finding useful lemmas, since existing lemmas may inspire the generation of similar lemmas tailored for the stuck proof state. \modify{\name{} applies both mechanisms rather than choosing between the two.} We elaborate on these two mechanisms below.

\textbf{Directly Propose New Lemmas.}
Proposing new lemmas requires an agent not only to formulate a statement that is both syntactically valid and helpful, but also to generate a correct proof of the statement. To address this, \name{} employs a two-stage process that decouples the generation of lemma statements from the generation of their proofs.

In the first stage, an LLM is instructed to generate statements of new lemmas. As shown in the prompt in Figure~\ref{prompt_lemma_discovery}, it is provided with existing theorems and lemmas in the working context and stuck proof state, so that it can generate semantically relevant lemma statements, while avoiding redundant lemmas that already exist. These generated statements are then checked by \coq{} for syntactic and semantic correctness. Any invalid statements are discarded.

In the second stage, \name{} attempts to generate a proof for each valid lemma statement, using the same procedure as the initial proof generation described in Section~\ref{sec:initial_proof}. The statement of a successfully proven lemma is added to the working context.

\begin{figure}[thb]
\footnotesize
\begin{tabular}{>{\raggedright\arraybackslash\tt}p{0.45\textwidth}}
\toprule
\headercolorlong
\textbf{INSTRUCTIONS} \\
We are proving a theorem in Coq but get stuck in a proof state, your task is to propose some new lemma statements to help solve it. You will be given the theorem statement and proof state, together with related definitions, existing theorem and lemmas. Your lemmas must state new, non-trivial properties. For each proposed lemma statement, explain why it is correct and useful. \\
\headercolorlong
\textbf{INPUT}\\
- Theorem Statement: \highlighttext{<theorem>} \\
- Proof State: \highlighttext{<proof state>} \\
- Definitions: \highlighttext{<definitions>} \\
- Existing Theorems and Lemmas: \highlighttext{<theorems and lemmas>} \\
\bottomrule
\end{tabular}

\caption{\label{prompt_lemma_discovery} Prompt template for proposing new lemmas.}
\end{figure}

\textbf{Adapt Existing Lemmas.}
Some existing theorems and lemmas, while not directly applicable, can be adapted to work in the current proof. For example, the existing lemma \texttt{n\_le\_m\_Sn\_le\_Sm} (Figure~\ref{example:context}, line 28) cannot be applied to the proof state in Figure~\ref{example:state_2}. However, a human developer might observe that its converse could be helpful and propose the new lemma \texttt{Sn\_le\_Sm\_n\_le\_m} shown in Figure~\ref{example:lemma}.

To refine a theorem or lemma selected by the decision-maker, \name{} provides an LLM with its statement and proof. The prompt, as shown in Figure~\ref{prompt_lemma_refine}, instructs the LLM to analyze how the lemma can be adapted to solve the proof goal. It then directs the model to generate both the refined statement and proof, encouraging it to reuse the original proof to simplify the proof generation task. Finally, \name{} validates each refined lemma by executing its statement and proof in \coq{}. Only those lemmas with both a well-formed statement and a correct proof are added to the working context.

After updating the working context, \name{} prompts an LLM to regenerate a proof. The prompt provides the original theorem statement, the updated working context, and the \proofstate{} from the prior proof failure. It also instructs the LLM to leverage the new lemmas in its proof attempt.

\begin{figure}[thb]
\footnotesize
\begin{tabular}{>{\raggedright\arraybackslash\tt}p{0.45\textwidth}}
\toprule
\headercolorlong
\textbf{INSTRUCTIONS}\\
We are proving a theorem in Coq but get stuck in a proof state, your task is to refine an existing lemma to help solve it. You will be given the theorem statement to prove, the stuck proof state, related definitions and the specific lemma to refine, including its proof. A set of existing theorems and lemmas are also provided. First analyze how to adapt this given lemma to help solve the proof goal, then produce the refined lemma statement together with its proof. Try to reuse the original proof of the given lemma where possible. \\
\headercolorlong
\textbf{INPUT}\\
- Theorem Statement: \highlighttext{<theorem>} \\
- Proof State: \highlighttext{<proof state>} \\
- Definitions: \highlighttext{<definitions>} \\
- Existing Theorems and Lemmas: \highlighttext{<theorems and lemmas>} \\
- Lemma to Refine: \highlighttext{<lemma and proof>} \\
\bottomrule
\end{tabular}

\caption{\label{prompt_lemma_refine} Prompt template for adapting existing lemmas.}
\vspace{-10pt}
\end{figure}

\subsection{Context Enrichment Strategy}
\label{sec:retrieval}
When this strategy is selected, \name{} searches for additional definitions, lemmas, theorems, and proofs from the global context to enrich its working context. Specifically, \modify{the decision-maker identifies a set of keywords based on the \proofstate{} and context, as described in Section~\ref{sec:decision_maker}. \name{} then performs a targeted search by matching these keywords against the statements of available theorems and lemmas.} For example, recognizing that the proof state in Figure~\ref{example:state_2} involves the relation \texttt{le} and function \texttt{leb}, the LLM may suggest these terms as keywords to find more relevant lemmas when the existing ones are deemed insufficient. Similarly, if \name{} has retrieved the proof of \texttt{leb\_complete}, it may identify the lemmas \texttt{n\_le\_m\_Sn\_le\_Sm} and \texttt{O\_le\_n} used in its proof, and generate their names as keywords if they are not in the working context. Subsequently, \name{} employs the same proof generation procedure used after lemma discovery described in Section~\ref{sec:lemma}, but supplies it with the enriched working context and instructs the LLM to leverage the newly retrieved context.

This design contrasts with prior work~\cite{thompson2025rango}, which selects theorems and lemmas by assessing the similarity between their statements and the entire proof state. The intuition is that the entire proof state may contain terms that are not critical for making progress and therefore using the entire proof state in a query can retrieve irrelevant theorems and lemmas or miss important ones. For example, in Figure~\ref{example:state_2}, the proof state contains common types like \texttt{nat} and \texttt{bool}. Including them in the query would lead the retrieval system to fetch broadly related but unhelpful lemmas about natural numbers or booleans, potentially missing a lemma \texttt{n\_le\_m\_Sn\_le\_Sm} particularly relevant to the inequality in the current goal. Section~\ref{sec:ablation} validates this design choice by comparing \name{} to \namefullstate{}, a variant of \name{} that utilizes all terms in the stuck proof state for searching instead of LLM-generated keywords. The result shows that using LLM-generated keywords indeed leads to more theorems to be proved (651 vs.~625).

\subsection{Regeneration Strategy}
The regeneration strategy can either repair simple errors in the previous proof, or explore a new proof approach when the previous one is deemed infeasible. When this strategy is selected, \name{} provides an LLM with the target theorem statement, the working context, and the \proofstate{} from the current proof failure. This strategy prompts the LLM to reason about the previous failure and then generate a new proof (Figure~\ref{prompt_regen}).

\begin{figure}[!t]
\footnotesize
\begin{tabular}{>{\raggedright\arraybackslash\tt}p{0.45\textwidth}}
\toprule
\headercolorlong
\textbf{INSTRUCTIONS}\\
I’m proving a theorem in Coq but got stuck in a proof state. You will be given the theorem statement to prove and the diagnostic information, including the stuck proof state, the erroneous tactic that causes the error, the error message from Coq and the partial proof preceding the erroneous tactic. Additional context, including related definitions, theorems and lemmas, and a similar theorem with its proof are also provided. First analyze how to prove the theorem based on the provided context, then write a correct proof. You can repair the proof if the error is simple, or generate a new proof if the prior proof is infeasible. \\

\headercolorlong
\textbf{INPUT}\\
- Theorem Statement: \highlighttext{<theorem>} \\
- Stuck Proof State: \highlighttext{<proof state>} \\
- Erroneous Tactic: \highlighttext{<erroneous tactic>} \\
- Error Message: \highlighttext{<error message>} \\
- Partial Proof: \highlighttext{<partial proof>} \\
- Definitions: \highlighttext{<definitions>} \\
- Theorems and Lemmas: \highlighttext{<theorems and lemmas>} \\
- Similar Proof: \highlighttext{<similar proof>} \\
\bottomrule
\end{tabular}

\caption{\label{prompt_regen} Prompt template for regeneration.}
\end{figure}

\section{Evaluation}

We investigate five research questions in our evaluation:

\begin{itemize}
\item \textbf{RQ1:} How effective is \name{} compared to state-of-the-art proof generation tools?
\item \textbf{RQ2:} How sensitive is \name{} to the underlying LLMs used?
\item \textbf{RQ3:} How much does each proof refinement strategy contribute to the overall effectiveness of \name{}?
\item \textbf{RQ4:} How does our LLM-guided decision-maker compare to other alternative designs?
\item \textbf{RQ5:} How does \name{}'s iteration limit impact its performance?

\end{itemize}

\subsection{Benchmarks}
We perform experiments using two different sets of benchmarks. The first is \coqstoq{}, a state-of-the-art benchmark based on 14 real-world \coq{} projects on GitHub~\cite{thompson2025rango}. We use its cut-off split, created to mitigate data leakage issues, consisting of 1,171 theorems.

We have also built a new set of benchmarks, called \bench{}, that represents a more realistic and challenging setting for automatic proof generation. The motivation is based on our observation that proof generation tasks from existing benchmarks are typically constructed by simply omitting the proof of the target theorem, leaving any helper lemmas and similar proofs available to the LLM. In practice, however, the relevant helper lemmas are introduced while constructing a proof, and similar proofs may be subsequently introduced after the commit of the proof. In other words, this simple masking-in-the-middle evaluation setting provides the LLM with important information that may not exist when the proof was actually written. Therefore, this setup may not fully reflect the capabilities of proof generation tools in real-world settings.

We constructed \bench{} by mining the commit histories of open-source \coq{} projects on GitHub. Specifically, we searched for projects using \coq{} as their primary language as of February 20, 2025, and ranked them by creation date, prioritizing newer projects to mitigate potential data leakage. We only included projects that can be built automatically with \textit{Makefile} or \textit{\_CoqProject}, resulting in a final set of \benchprojects{} projects. These projects span a range of verification domains, including logic, data structures, programming languages, and mathematics. We collected all commits of each selected project to get 2,225 commits in total. Next, we compare each commit against its parent to identify any newly introduced or modified theorems.

For each theorem, we masked its proof body as well as the other theorems and lemmas introduced in the commit. In this way, we make sure that LLMs cannot access any helper lemmas or theorems introduced during the construction of the proof. As a result, LLMs need to propose these lemmas or theorems as human developers did when writing the proof. In addition, any theorems and lemmas introduced in future commits are also not available to LLMs. Only the previously committed lemmas and theorems are available to the LLMs, mimicking how proofs are actually developed in practice. \modify{On average, 60.34 same-commit lemmas and theorems are masked per task. Moreover, 23.2\% of the human-written proofs use at least one declaration introduced in the same commit, indicating that the masked context is relevant to proof construction.} To prevent a few large projects from dominating the benchmark and ensure diversity, we randomly sampled at most 200 theorems from each project to form the final benchmark. In the end, we collected \benchtheorems{} theorems. \modify{The average number of tactics in the human-written proofs is \benchavgsteps{}, and the median is \benchmediansteps{}.}

\subsection{Comparison Baselines}
We compare \name{} against three state-of-the-art LLM-based \coq{} proof generation tools. We select \cobblestone{}~\cite{kasibatla2025cobblestone} and \rango{}~\cite{thompson2025rango} as they achieved state-of-the-art performance on the most recent \coqstoq{} benchmark. We also include \palm{}~\cite{lu2024proof} which employs hard-coded rules to select a refinement strategy. Since \palm{} does not query the LLM during refinement, we build two additional baselines, \selfrefine{} and \selfrefinerag{}, that refine proofs with iterative LLM queries. Each baseline is described below.

\begin{itemize}
\item \textbf{\cobblestone{}}~\cite{kasibatla2025cobblestone} adopts a divide-and-conquer strategy. It prompts an LLM to generate whole proofs, preserves the successful sub-proofs, and queries the LLM to resolve remaining unproven subgoals.

\item \textbf{\rango{}}~\cite{thompson2025rango} fine-tunes the DeepSeek-Coder 1.3B~\cite{guo2024deepseek} for step-by-step proof generation. At each proof step, it retrieves helper lemmas and existing proofs based on their similarity to the current proof state.

\item \textbf{\palm{}}~\cite{lu2024proof} employs hard-coded rules to select repair heuristics that fix common errors based on error types. If the heuristics fail, it falls back to a backtracking strategy to incrementally regenerate previous proof steps with CoqHammer~\cite{coqhammer}.

\item \textbf{\selfrefine{}}~\cite{madaan2023self} adopts a simple iterative repair loop. After the initial proof failure, it repeatedly prompts the LLM to generate a new proof based on the \proofstate{} from the previous failure, without a working context, further lemma discovery or context enrichment.

\item \textbf{\selfrefinerag{}} enhances \selfrefine{} by initializing the working context before entering the iterative repair loop. In each iteration, it instructs the LLM to explicitly analyze the \proofstate{} and working context, then regenerate a proof. However, it does not perform further adaptive lemma discovery or context enrichment.
\end{itemize}

\subsection{Experimental Setup}
All experiments are conducted on a server with an AMD EPYC 7313 CPU, an NVIDIA A5500 GPU, and \SI{128}{GB} memory, running 64-bit Ubuntu 22.04 LTS.

\modify{To enable a fair comparison with \cobblestone{} and \palm{}, we allow \name{}, \selfrefine{}, and \selfrefinerag{} to use \coqhammer{}. We first invoke \coqhammer{} on the initial proof state, and once more when a tactic fails. In the latter case, if \coqhammer{} fails to solve the goal, the proof attempt is considered unsuccessful. We have not modified \rango{} to use \coqhammer{}, as doing so would require significant changes to its source code.}

In all experiments, we set the temperature to 0 to mitigate randomness. In all experiments except the one in RQ2, we use GPT-4o-2024-08-06~\cite{gpt4o} as the underlying LLM for \name{} and the non-\rango{} comparison baselines. \rango{} uses a finetuned LLM for step-by-step proof generation, and our experiments use the finetuned DeepSeek-Coder model from its original implementation. We choose not to finetune GPT-4o for \rango{} due to its high cost. In RQ2, we further experiment with five additional popular LLMs with varying parameter sizes as the backbone of \name{}. These models include \gptfive{}~\cite{gpt52}, \gptfouromini{}~\cite{gpt4omini}, \claude{}~\cite{claude37}, \deepseek{}~\cite{liu2024deepseek}, and \llama{}~\cite{llama4}. Due to the high computational and token costs, we conduct the RQ2 experiment on 500 randomly sampled theorems from \bench{}.

In RQ3, we conduct an ablation study to measure the contribution of \name{}'s proof refinement strategies. Specifically, we created three variants of \name{}: (1) \namenolemma{}, which disables the lemma discovery strategy; (2) \namenorag{}, which disables the context enrichment strategy; and (3) \namebase{}, which retains only the regeneration strategy. When a strategy is disabled, we modify the prompt for the decision-making LLM accordingly to prevent its selection. Furthermore, to validate our keywords-based searching within the context enrichment strategy, we implemented a variant called \namefullstate{} that utilizes all terms in the stuck proof state for searching instead of LLM-generated keywords.

In RQ4, we investigate alternative designs of the decision-maker used by \name{}, experimenting with deep learning-based, rule-based, random, and sequential approaches. Section~\ref{sec:decision_strategy} describes each decision-maker in more detail.

In all experiments except the one in RQ5, we set the refinement iteration limit to 3 for \name{} and the baselines. RQ5 investigates how the limit on refinement iterations impacts \name{}'s performance.

\subsection{RQ1: Overall Effectiveness of \name{}}
Table~\ref{table_bench} presents the number and percentage of theorems proven by each approach on both the existing \coqstoq{} and \bench{}. Overall, \name{} significantly outperforms all baselines on both benchmarks. On \coqstoq{}, \name{} proves \oursoutcobblestoneoncoqstoq{}, \oursoutpalmoncoqstoq{}, and \oursoutrangooncoqstoq{} more theorems than \cobblestone{}, \palm{} and \rango{}, respectively. On \bench{}, which simulates a more realistic development scenario with limited context, \name{}'s advantage is maintained. It outperforms the baselines by \oursoutcobblestoneonbench{}, \oursoutpalmonbench{}, and \oursoutrangoonbench{}, respectively. Moreover, it surpasses the strongest baseline, \selfrefinerag{}, by \oursoutselfrefineragonbench{}. These results demonstrate the effectiveness of \name{} in integrating and dynamically selecting different refinement strategies in proof generation compared to state-of-the-art methods.

\begin{table}[t]
\begin{center}
\caption{Theorems proven by each approach.}
\resizebox{0.48\textwidth}{!}{
\begin{tabular}{lccc}
\toprule
\multirow{2}{*}{\textbf{Approach}} & \multicolumn{2}{c}{\textbf{\# of Theorems Proven}} & \multirow{2}{*}{\textbf{Avg. Cost (\$)}} \\
\cmidrule{2-3}
& \textbf{\bench{}} & \textbf{\coqstoq{}} \\
\midrule
\cobblestone{} & \cobblestoneonbench{} (\cobblestoneonbenchp{}) & \cobblestoneoncoqstoq{} (\cobblestoneoncoqstoqp{}) & 0.154 \\
\rango{} & \rangoonbench{} (\rangoonbenchp{}) & \rangooncoqstoq{} (\rangooncoqstoqp{}) & - \\
\palm{} & \palmonbench{} (\palmonbenchp{}) & \palmoncoqstoq{} (\palmoncoqstoqp{}) & 0.012\\
\selfrefine{} & \selfrefineonbench{} (\selfrefineonbenchp{}) & \selfrefineoncoqstoq{} (\selfrefineoncoqstoqp{}) & 0.051 \\
\selfrefinerag{} & \selfrefineragonbench{} (\selfrefineragonbenchp{}) & \selfrefineragoncoqstoq{} (\selfrefineragoncoqstoqp{}) & 0.058 \\
\name{} & \oursonbench{} (\oursonbenchp{}) & \oursoncoqstoq{} (\oursoncoqstoqp{}) & 0.085 \\

\bottomrule
\end{tabular}
}
\label{table_bench}
\end{center}
\end{table}

\modify{We can also compare the sets of theorems uniquely proved by either \name{} or one of our baselines. On \bench{}, \name{} proves \oursuniquebench{} (\oursuniquebenchp{}) theorems that no baseline proves, while \baselineuniquebench{} (\baselineuniquebenchp{}) theorems are proved by at least one baseline but not by \name{}. On \coqstoq{}, the corresponding counts are \oursuniquecoqstoq{} (\oursuniquecoqstoqp{}) and \baselineuniquecoqstoq{} (\baselineuniquecoqstoqp{}), respectively. These results show that \name{} is complementary to the baselines.}

\begin{figure}[t]
\includegraphics[width=0.5\textwidth]{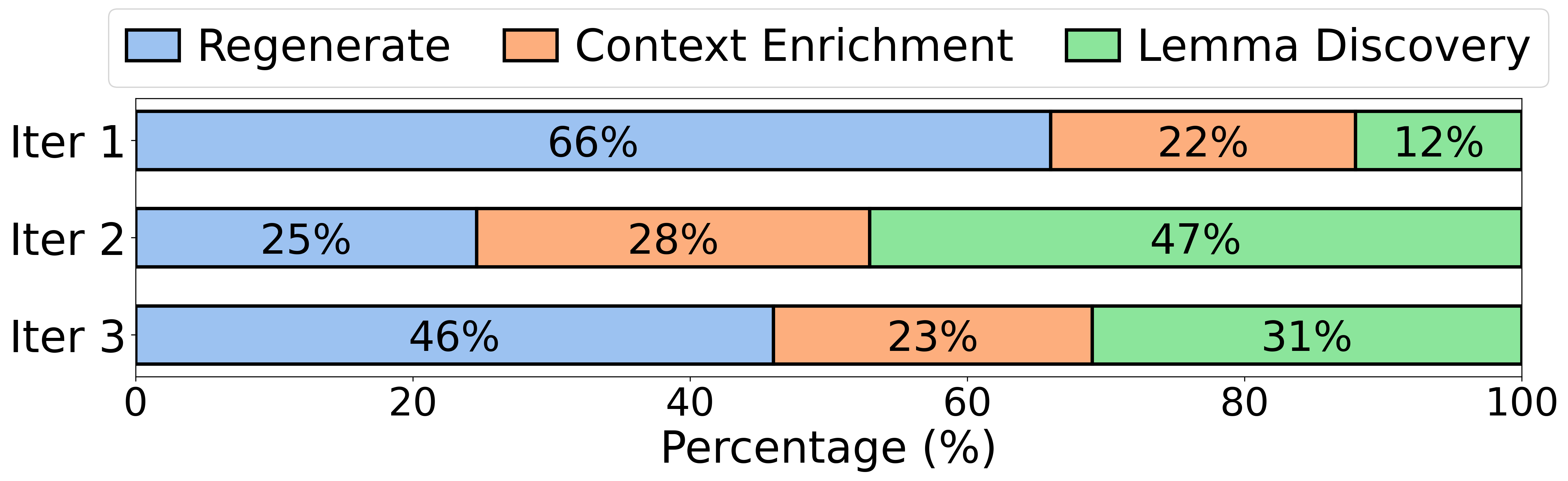}
\caption{\label{eva:choice_iter} Percentage of strategies selected in each iteration.}
\end{figure}

To better understand the refinement strategies selected by \name{}, we analyzed the percentage of tasks where \name{} selected each strategy in each refinement iteration, as shown in Figure~\ref{eva:choice_iter}. The result shows that the LLM decision-maker in \name{} selects different strategies throughout the refinement loop, instead of adhering to a single fixed approach. This indicates that each strategy contributes to solving different proof challenges.

\noindent{\bf\em Cost Analysis.} On average, \name{} costs \$0.085 USD per theorem, due to the multiple LLM calls required during iterative refinement. \selfrefine{} and \selfrefinerag{} cost \$0.051 USD and \$0.058 USD per theorem, respectively, as they also repeatedly regenerate proofs after failures. \cobblestone{} has the highest cost at \$0.154 USD per theorem. This is expected based on its divide-and-conquer strategy which prompts the LLM to discharge unproven subgoals. \palm{} only costs \$0.012 USD per theorem, since it only prompts an LLM to generate the initial proof, and relies solely on symbolic methods to repair any proof failures. \rango{} incurs the minimal cost, since it runs a fine-tuned local model rather than querying a closed-source API. Overall, these results highlight a trade-off between cost and performance. \name{} achieves the best performance at a moderate cost.

\subsection{RQ2: Sensitivity of \name{} to LLM Choices}

Table~\ref{table_llms} presents the number of theorems proven by \name{} using each underlying LLM. We observe that \name{} performs the best with \gptfive{}, successfully proving \oursfive{} theorems. The default model used in RQ1, \gptfouro{}, also performs strongly, proving \oursfouro{} theorems. Its performance remains comparable when using other large models such as \claude{}, \deepseek{}, and \llama{}, proving \oursclaude{}, \oursdeepseek{}, and \oursllama{} theorems, respectively. As expected, \name{} proves 19.15\% fewer theorems when using \gptfouromini{}, which is significantly smaller (estimated to have \SI{8}{B} parameters) compared to \gptfouro{} (estimated to have \SI{200}{B} parameters). This is a reasonable reduction given the limited capability of smaller models. However, \name{}'s performance with \gptfouromini{} is still comparable to state-of-the-art baselines like \palm{}, which uses the more powerful \gptfouro{} model in RQ1.

\subsection{RQ3: Contribution of Refinement Strategies}
\label{sec:ablation}
Table~\ref{table_abl} presents the performance of these variants compared to the complete \name{} framework on the \bench{} benchmark. We observe that the full \name{} proves \oursoutnolemma{} and \oursoutnorag{} more theorems than \namenolemma{} and \namenorag{}, respectively. These results indicate that both lemma discovery and context enrichment contribute to the overall effectiveness of \name{}, and lemma discovery is particularly critical. Furthermore, \namebase{} performs comparably to \selfrefine{}, which is expected because \namebase{} functions as a standard iterative refinement loop. Finally, replacing generated keywords with the full proof state for searching reduces the success rate by 4.00\% and increases token consumption by 7.44\%, confirming the efficiency of our keyword-based approach.

\begin{table}[t]
\centering
    \caption{Theorems proven by \name{} using different LLMs \modify{on 500 theorems sampled from \bench{}.}}
    \label{table_llms}
    \begin{tabular}{lc}
    \toprule
    \textbf{Approach} & \textbf{\makecell{\# of Theorems Proven}} \\ \midrule
    \gptfive{} & \oursfive{} (\oursfivep{}) \\
    \gptfouro{} & \oursfouro{} (\oursfourop{}) \\
    \gptfouromini{} & \oursfouromini{} (\oursfourominip{}) \\
    \claude{} & \oursclaude{} (\oursclaudep{}) \\
    \deepseek{} & \oursdeepseek{} (\oursdeepseekp{}) \\
    \llama{} & \oursllama{} (\oursllamap{}) \\
    \bottomrule
    \end{tabular}
\end{table}

\begin{table}[t]
    \centering
    \caption{Effectiveness of refinement strategies on \bench{}.}
    \label{table_abl}
    \begin{tabular}{lc}
    \toprule
      \textbf{Variant} & \textbf{\makecell{\# of Theorems Proven}} \\ \midrule
      \namenolemma{} & \oursnolemma{} (\oursnolemmap{}) \\
    \namenorag{} & \oursnorag{} (\oursnoragp{}) \\
    \namebase{} & \oursbase{} (\oursbasep{}) \\
    \namefullstate{} & \oursfullstate{} (\oursfullstatep{}) \\
    \name{} & \oursonbench{}{} (\oursonbenchp{}) \\
    \bottomrule
    \end{tabular}
\end{table}

\subsection{RQ4: Comparing Decision-Maker Designs}
\label{sec:decision_strategy}

While \name{} leverages the reasoning capabilities of LLMs for adaptive strategy selection based on reflection over the current proof progress and available context, we also investigate alternative decision-making mechanisms to understand the trade-offs.

\vspace{1mm}
\noindent\textbf{Learning-Based Decision Making.} Another mechanism we explored is a learning-based approach, which employs a machine learning (ML) model to predict the strategy a developer would use in a given \proofstate{}. Training such a model requires a labeled dataset of the \proofstate{} and the corresponding strategies taken by human developers, which is not available. To address this, we collect a training dataset of 18,628 examples from the human-written proofs in \bench{}, and evaluate the model using ten-fold cross-validation.

To generate each data point in the training dataset, we first simulate the working context at each step of a human-written proof by using the same retrieval method described in Section~\ref{sec:context_init}. We then extract the following features to characterize the \proofstate{}:
\begin{enumerate}
    \item The number of unique definitions appearing in the proof goal.
    \item The number of unique definitions appearing in the local hypotheses.
    \item The count of definitions that appear in the goal but not in the hypotheses.
    \item The count of definitions that appear in the goal but not in existing lemmas in the working context.
    \item The similarity score of the retrieved similar theorem.
    \item The number of lemmas used in the similar proof that are not present in the working context.
\end{enumerate}
We select features 1 and 2 to reflect proof state complexity. Features 3--5 measure how many definitions in the goal lack supporting lemmas in the working context. Feature 6 assesses if potentially useful information in a similar theorem has not been incorporated.

Given the extracted features, the ML model predicts which refinement strategy to take at the given proof state, which is a typical classification problem. The ground-truth labels are derived from the corresponding human-written tactic at the proof state: if the tactic applies a new helper lemma introduced in the same commit, it is labeled as \textit{Lemma Discovery}; if it applies a lemma that existed prior to the current commit, it is labeled as \textit{Context Enrichment};  otherwise, it is labeled as \textit{Regeneration}.

We conduct a preliminary experiment using ten-fold cross validation to compare several ML models and select the best one to be compared with other types of decision-makers. We evaluate traditional models, including Support Vector Machine (SVM)~\cite{hearst1998support}, Decision Tree~\cite{safavian1991survey}, Random Forest~\cite{breiman2001random}, and XGBoost~\cite{chen2016xgboost}. In addition, we build a Deep Neural Network (DNN)~\cite{lecun2015deep} with three fully-connected layers, using ReLU~\cite{nair2010rectified} as the activation function and a dropout rate of 0.2. We train this DNN for 50 epochs using the Adam optimizer~\cite{kingma2017adam}, with a learning rate of 0.001. The average accuracy and F1 score of each model is shown in Table~\ref{table_ml}. We ultimately select DNN as it achieves the highest average accuracy and F1 score.

\begin{table}[t]
    \centering
    \caption{Average accuracy and F1 scores for each model.}
    \label{table_ml}

    \small
    \begin{tabular}{lcc}
    \toprule
    \textbf{Model} & \textbf{Acc.(\%)} & \textbf{F1} \\
    \midrule
    SVM & 41 & 0.48 \\
    Decision Tree & 25 & 0.29 \\
    Random Forest & 41 & 0.47 \\
    XGBoost & 71 & 0.74 \\
    DNN & \textbf{82} & \textbf{0.78} \\
    \bottomrule
    \end{tabular}
\end{table}

\begin{table}[t]
    \centering
    \caption{Performance and cost comparison of \name{} with different decision-makers on \bench{}.}
    \label{table_decision}
    \small
    \begin{tabular}{lccc}
    \toprule
    \textbf{Approach} & \textbf{\makecell{\# of Theorems\\Proven}} & \textbf{\# Iters} & \textbf{\makecell{\# of Tokens\\(Input / Output)}} \\
    \midrule
    \namemldecision{} & 595 (34.59\%) & 1.24 & 9801 / 3157 \\
    \nameruledecision{} & 647 (37.62\%) & 1.37 & 24244 / 6109 \\
    \namerandomdecision{} & 605 (35.17\%) & 1.61 & 12568 / 3476 \\
    \nameseqdecision{} & 613 (35.64\%) & 1.58 & 11376 / 3296 \\
    \name{} & 651 (37.85\%) & 1.35 & 16348 / 4383 \\
    \bottomrule
    \end{tabular}
\end{table}

\vspace{1mm}
\noindent\textbf{Rule-Based Decision Making.} One alternative is a rule-based method, which employs heuristics to select a strategy. First, it checks whether the retrieved similar proof uses lemmas that are missing from the current working context. If so, it prioritizes context enrichment. Next, it analyzes the definitions (e.g., types and functions) in the current proof state. If the proof state involves definitions that are not present in any lemmas in the working context, it selects lemma discovery. The intuition is that formal proofs rely on lemmas to specify the properties of definitions. A definition in the goal without corresponding lemmas indicates an absence of established properties and therefore requires lemma discovery. If neither of these two conditions is met, it defaults to Regeneration.

\vspace{1mm}
\noindent\textbf{Random Decision Making.} We also implement a baseline decision-maker that randomly selects one of the refinement strategies in each iteration. This is expected to be the weakest baseline, as it does not incorporate any reasoning or contextual information when choosing a strategy.

\vspace{1mm}
\noindent\textbf{Sequential Decision Making.} We further evaluate \nameseqdecision{}, a deterministic baseline that cycles through strategies in a fixed, cost-ascending order: (1) Regeneration, (2) Context Enrichment, and (3) Lemma Discovery.

\vspace{1mm}
\noindent\textbf{Results.} Table~\ref{table_decision} compares the success rates and average costs of different decision-makers in terms of refinement iterations and token consumption. Overall, the results indicate that the LLM-guided decision-maker achieves the best balance of performance and efficiency. Although the rule-based decision-maker achieves a comparable success rate, it is significantly less efficient, consuming nearly 50\% more tokens on average. This inefficiency stems from its simple heuristics, which often unnecessarily select the resource-intensive lemma discovery strategy.

The importance of intelligent strategy selection is further highlighted by the random baseline, which proves 7.07\% fewer theorems than the LLM decision-maker and requires, on average, 19.26\% more iterations to find a proof. The learning-based decision-maker using DNN performs even worse than the random baseline. We attribute its underperformance to two key challenges: first, our quantitative features may not fully capture the complex semantics of a proof state, and second, the scarcity of training data, especially for lemma discovery, makes the model overly conservative.

\begin{figure}[t]
\includegraphics[width=0.48\textwidth]{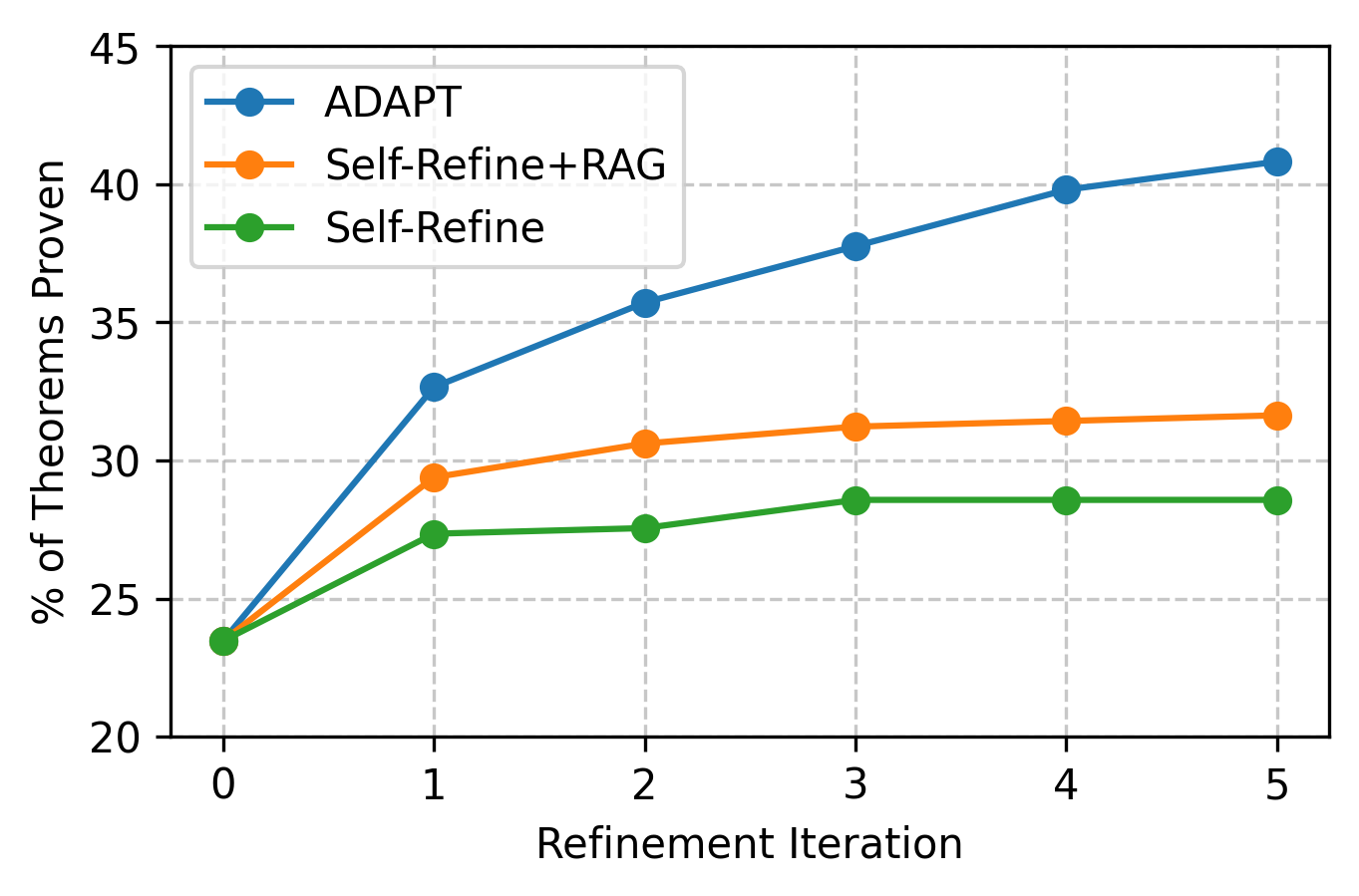}
\caption{\label{eva:succ_by_iters} Percentage of theorems proven by refinement iterations.}
\end{figure}

\subsection{RQ5: Impact of Refinement Iteration Limit}
Figure~\ref{eva:succ_by_iters} shows the cumulative percentage of theorems successfully proven on \bench{} as the iteration count increases. We observe that while all three approaches benefit from more iterations, the success rates for \selfrefine{} and \selfrefinerag{} begin to saturate after 3 attempts. In contrast, \name{} continues to improve its performance more steadily across all 5 iterations. This suggests that \name{}'s ability to introduce powerful new lemmas via lemma discovery and context enrichment provides it with more opportunities to make progress on challenging theorems where simpler iterative repair methods have stalled.

\begin{figure}
\centering
\lstset{
    style=coq,
    basicstyle=\ttfamily\footnotesize,
    breaklines=true,
    frame=none,
    xleftmargin=0pt,
    aboveskip=0pt,
    belowskip=0pt
}
\begin{subfigure}[htb]{\linewidth}
\begin{lstlisting}
(*@\textcolor{gray}{(* Required existing lemma *)}@*)
Lemma Ordered_iff_Ordered' t :
  Ordered t <-> Ordered' t.

(*@\textcolor{gray}{(* Theorem statement and human-written proof *)}@*)
Lemma rotate_right_Ordered_imp v l r :
  Ordered (rotate_right v l r) -> Ordered (Node v l r).
Proof.
  destruct l as [|lv ll lr]; eauto 2.
  rewrite !Ordered_iff_Ordered'.
  ...
\end{lstlisting}
\caption{Incorrect strategy selection: \modify{\name{} selects regeneration in all iterations and never retrieves the required existing lemma~\cite{case1}.}}

\label{case:1}
\end{subfigure}

\begin{subfigure}[htb]{\linewidth}
\vspace{.2cm}
\begin{lstlisting}
(*@\textcolor{gray}{(* Required lemma statement correctly proposed by \name{} *)}@*)
Lemma valt_dec : forall T1 n n1 v1,
  val_type n v1 T1 -> n1 <= n -> val_type n1 v1 T1.

(*@\textcolor{gray}{(* Theorem statement and human-written proof *)}@*)
Lemma envt_dec : forall n n1 H G,
  env_type n H G -> n1 <= n -> env_type n1 H G.
Proof.
  ...
  eapply valt_dec; eauto.
Qed.
\end{lstlisting}
\caption{Ineffective strategy execution: \modify{\name{} correctly proposes the required helper lemma but fails to prove it~\cite{case2}.}}
\label{case:2}
\end{subfigure}

\begin{subfigure}[htb]{\linewidth}
\vspace{.2cm}
\begin{lstlisting}
(*@\textcolor{gray}{(* Required existing lemma correctly retrieved by \name{} *)}@*)
Lemma const_unfold : forall T (x : T),
  const x = Cons x (const x).

(*@\textcolor{gray}{(* Theorem statement *)}@*)
Lemma L_as_K : forall n, L n = K n << 0.
Proof.
  (*@\textcolor{gray}{(* Human-written proof *)}@*)
  destruct n as [| n].
  - apply const_unfold.
  ...

  (*@\textcolor{gray}{(* LLM-generated proof *)}@*)
  destruct n.
  - rewrite const_unfold.
  ...
\end{lstlisting}
\caption{Proof generation failure: \modify{the required lemma is available, but \name{} applies it using an invalid tactic~\cite{case3}.}}
\label{case:3}
\end{subfigure}
\caption{Representative failure examples from \bench{}.}
\label{fig:coq_cases}
\end{figure}

\subsection{Failure Analysis}
To understand the limitations of our approach, we manually analyzed 100 randomly sampled theorems from the \bench{} benchmark that \name{} failed to prove. Our analysis identified three primary failure modes:

\vspace{1mm}
\noindent\textbf{1. Incorrect Strategic Decisions (39\%).} A primary cause of failure is the decision-maker's inability to identify the correct strategy. Specifically, in 25\% of cases, the LLM failed to select context enrichment in any refinement iteration, despite the need for existing lemmas. For instance, Figure~\ref{case:1} shows an example where the LLM repeatedly selected regeneration, failing to find the necessary lemma. Similarly, in 14\% of cases, it never attempted lemma discovery, though new helper lemmas were required.

\vspace{1mm}
\noindent\textbf{2. Ineffective Strategy Execution (32\%).} Even when the correct strategy is selected, the execution modules may still fail. In 17\% of cases, \textit{Context Enrichment} failed to select required lemmas because the LLM generated ineffective search keywords. In 15\% of cases, the lemma discovery strategy only proposed trivial lemmas or failed to prove the proposed lemmas. For instance, Figure~\ref{case:2} illustrates an example where the lemma discovery strategy correctly proposed the statement of the required lemma, but failed to prove it.

\vspace{1mm}
\noindent\textbf{3. Proof Generation Failures (29\%).} Despite correct strategic decisions and execution, the LLM may still fail to generate the correct proof script. In 9\% of cases, the LLM failed to utilize the theorems or synthesized lemmas, although they were provided in the prompt. The remaining 20\% of failures were caused by incorrect tactic applications or syntax errors. For instance, Figure~\ref{case:3} illustrates an example where the correct lemma (\texttt{const\_unfold}) is provided in the prompt, but the LLM mistakenly tried to apply it to the goal using the \texttt{rewrite} tactic instead of the \texttt{apply} tactic.

\section{Discussion}
\subsection{Threats to Validity}
\noindent \textbf{Internal Validity.} The main threat to internal validity arises from the inherent randomness of LLMs. To mitigate this threat, we set the temperature of all evaluated LLMs to 0 for more deterministic outputs. Furthermore, we conducted evaluations across two benchmarks and with six LLMs of different parameter sizes. The consistent results observed across these varied settings help to reduce this threat.

\vspace{1mm}
\noindent \textbf{External Validity.} The threat to external validity concerns the generalizability of our experimental results. To mitigate this threat, we conducted experiments on two benchmarks that include various real-world projects covering multiple domains. However, while the principles of \name{}'s approach are applicable to other interactive proof assistants such as Isabelle and Lean, our current implementation and evaluation are confined solely to the \coq{} ecosystem. Extending \name{} to and evaluating it within these other ITPs remains for future work.

\vspace{1mm}
\noindent \textbf{Construct Validity.} A potential threat to construct validity is that we measure the effectiveness of a decision-maker with the overall success rate, which is an indirect metric. This is because a good strategic choice does not guarantee a successful proof. For example, the decision-maker could select a suitable strategy, such as lemma discovery, but the execution of this strategy may fail, leading to an overall proof failure that does not reflect a poor decision.

\subsection{Limitations and Future Work}
The framework of \name{} can be expanded by integrating more refinement strategies, such as proof repair models fine-tuned on ITP feedback, similar to the approach of Baldur~\cite{first2023baldur}. Our decision-maker primarily leverages LLM reasoning, with initial explorations into rule-based and machine learning approaches. Future research could investigate more sophisticated methods. For instance, this could involve creating more nuanced heuristics for rule-based methods or incorporating richer features for the machine learning models.
Currently, \name{} relies solely on LLMs to propose lemma statements, but the quality of these LLM-generated statements can vary; some are trivial, syntactically flawed, or unprovable. Future work could enhance the lemma discovery strategy by developing advanced prompting techniques or integrating symbolic methods.

\section{Related Work}
\label{sec:related_work}

\vspace{1mm}
\noindent \textbf{Proof Refinement.}
Recent work in proof refinement has focused on combining a vanilla LLM proof generation model with a standalone repair procedure when errors occur. These approaches typically generate an initial proof attempt with an LLM and then employ various repair strategies that leverage proof error information. For instance, Baldur~\cite{first2023baldur} fine-tunes a second LLM specifically for proof repair, regenerating a new proof from both the incorrect proof and the \coq{} error message. SAFE~\cite{chen2024automated} fine-tunes an LLM for both proof generation and refinement.
PALM~\cite{lu2024proof} employs hard-coded rules to select bug-fixing heuristics based on error types. If the heuristics fail, it falls back to a backtracking strategy that incrementally reverts proof steps and leverages automated theorem provers such as CoqHammer~\cite{coqhammer} to regenerate those steps.
AutoVerus~\cite{yang2025autoverus} automates the verification of Rust programs using Verus~\cite{lattuada2023verus} by refining LLM-generated proofs based on the specific error type. \cobblestone{}~\cite{kasibatla2025cobblestone} employs a divide-and-conquer strategy. It samples whole proofs from the LLM and uses additional LLM calls to resolve failed subgoals while preserving successful proof steps. A common limitation of these methods lies in their rule-based decision-making mechanisms. In contrast, \name{} adopts an LLM-guided decision-maker to dynamically select suitable proof refinement strategies based on the proof failure and related context.

\vspace{1mm}
\noindent \textbf{Retrieval-Augmented Proof Generation.}
Another line of work enhances automated proof generation by integrating retrieval mechanisms for relevant context. A common approach is to retrieve premises like definitions and lemmas. For instance, ReProver~\cite{yang2024leandojo} fine-tunes a retrieval model to find relevant premises given a proof state. Magnushammer~\cite{mikula2023magnushammer} improves the retrieval process by training a reranking model to prioritize the most relevant premises. Furthermore, Rango~\cite{thompson2025rango} retrieves both relevant proofs and lemmas via textual similarity to the current proof state, providing richer guidance for the proof search. While these methods focus on enhancing proof generation with retrieval methods, \name{} integrates multiple strategies and selects from them adaptively. Furthermore, \name{} leverages pretrained LLMs, whereas these approaches often require significant training or fine-tuning.

\vspace{1mm}
\noindent \textbf{Lemma Discovery.}
Lemma discovery facilitates automated proof generation by proposing helper lemmas aimed at simplifying complex proofs. These techniques range from neuro-symbolic approaches to data-driven methods. For instance, Lemmanaid~\cite{alhessi2025lemmanaid} employs a neuro-symbolic pipeline where a fine-tuned model generates lemma templates with symbolic holes that a symbolic engine then instantiates with concrete variables. Lfind~\cite{sivaraman2022data} offers a data-driven alternative, framing lemma discovery as a synthesis task that generalizes the proof goal into a sketch and then finds instantiations. These dedicated lemma discovery techniques are complementary to our work. Integrating them into \name{}'s adaptive framework to improve the quality of proposed lemmas remains a promising avenue for future research.

LegoProver~\cite{wang2023lego} uses an LLM to propose and prove new lemmas. The verified lemmas are stored in a library and retrieved through embedding-based search. It also refines existing lemmas to improve their reusability. LemmaNet~\cite{zhao2026lemma} focuses on lemma discovery for proving verification conditions. It first analyzes the source code and specifications to synthesize and prove helper lemmas that are relevant to program semantics. It then adapts these lemmas to the verification condition using feedback from the proof assistant. Quarry~\cite{zhang2026planning} decomposes a proof by prompting an LLM to propose helper lemmas. It then ranks the proposed lemmas based on a difficulty model, and recursively proves each lemma. While these tools use lemma discovery as their main refinement mechanism, \name{} selects from multiple refinement strategies dynamically.



\vspace{1mm}
\noindent \textbf{Agentic Theorem Proving.}
\modify{COPRA~\cite{thakur2023context} leverages the in-context learning capabilities of LLMs to perform a backtracking search. At each step, it prompts the LLM with the current proof state, previous failures, and retrieved lemmas to generate the next tactic. Failed or non-progressing tactics are recorded and avoided in subsequent attempts. It backtracks to the preceding state after repeated unsuccessful attempts at a proof state. RocqStar~\cite{kozyrev2025rocqstar} employs a multi-agent framework. It first uses multi-agent debate to generate and rank candidate proof plans, and then assigns an executor agent to execute the selected plan. If execution repeatedly fails, a critic model analyzes the current proof state and context, after which a replanner model revises the plan for another execution attempt. AutoRocq~\cite{autorocq} represents the current proof as a tree and prompts the LLM to generate one tactic each iteration. Upon failures, it first prompts the LLM to regenerate the tactic, and prompts the LLM to generate a query to search for additional lemmas when an error persists after several regenerations. APOLLO~\cite{ospanov2026apollo} is an agentic framework for repairing Lean proofs. It uses compiler feedback to fix  syntax errors, localize failing sub-lemmas, apply automated solvers, and invoke an LLM on the remaining goals. All these systems employ a fixed workflow, while \name{} employs a decision-maker to dynamically select refinement strategies based on the \proofstate{} and current working context.}

\section{Conclusion}
In this paper, we propose \name{}, an adaptive proof refinement
framework. \name{} integrates multiple refinement strategies,
including lemma discovery, context enrichment and regeneration. Upon
proof failures, \name{} employs an LLM-based decision-maker to select
an appropriate strategy based on the \proofstate{} and available
context. We also introduce \bench{}, a new benchmark that better
simulates more challenging real-world development processes. Our
experiments show that \name{} significantly outperforms
state-of-the-art baselines on both \bench{} and the existing
\coqstoq{} benchmark. Further analysis demonstrates \name{}'s
generalizability on six different LLMs, and measures the contribution of its refinement strategies. We also compare the trade-offs of alternative decision-maker designs.

\section{Data Availability Statement}
A replication artifact is publicly available: \url{https://doi.org/10.5281/zenodo.21708212}

\begin{acks}
We thank the anonymous reviewers for their valuable suggestions and feedback. This work is supported in part by NSF grants CCF-2340408, CCF-2321680, and ITE-2333736.
\end{acks}

\bibliographystyle{ACM-Reference-Format}
\bibliography{ref}

\end{document}